\documentclass[aps,twocolumn,showpacs,showkeys,preprintnumbers,amsmath,amssymb]{revtex4}

\usepackage{txfonts}
\usepackage{dcolumn}
\usepackage{mathrsfs}% Align table columns on decimal point
\usepackage{bm}% bold math
\usepackage{amsmath,amssymb,epsfig,float}

\begin{document}

\title{ Dependence of entanglement on initial states \\ under amplitude damping channel in non-inertial frames }

\author{Wenpin Zhang, Junfeng Deng and Jiliang
Jing\footnote{Corresponding author, Email: jljing@hunnu.edu.cn}}
\affiliation{ Department of Physics,
and Key Laboratory of Low-dimensional Quantum Structures
\\ and Quantum
Control of Ministry of Education, \\ Hunan Normal University,
Changsha, Hunan 410081, P.  R.  China}

% \baselineskip=0.65 cm

%\vspace*{0.2cm}
\begin{abstract}
%\vspace*{0.2cm}

Under amplitude damping channel, the dependence
of the entanglement on the initial states $|\Theta\rangle_{1}$ and
$|\Theta\rangle_{2}$, which reduce to four orthogonal Bell states if
we take the parameter of states $\alpha=\pm 1/\sqrt{2}$ are investigated. We find
that the entanglements for different initial states will decay along
different curves even with the same acceleration and parameter of
the states. We note that, in an inertial frame, the sudden death of
the entanglement for  $|\Theta\rangle_{1}$ will occur if
$\alpha>1/\sqrt{2}$, while it will not take place for
$|\Theta\rangle_{2}$ for any $\alpha$. We also show that the
possible range of the sudden death of the entanglement for
$|\Theta\rangle_{1}$ is larger than that  for  $|\Theta\rangle_{2}$.
There exist two groups of Bell state here we can't distinguish only by concurrence.

\end{abstract}

\vspace*{0.5cm}


\maketitle

 \section{introduction}

In the theory of quantum information, entanglement, a very subtle
phenomenon, has been investigated many years since it was first
brought to light by Einstein, Podolsky and Rosen \cite{a}, and by
Schr\"{o}dinger \cite{b,c}. It took about 30 years to distinguish it
from classical physics by Bell \cite{d}, and it was also found that
the entanglement plays a key role in quantum computation algorithms
\cite{e}. To the best of our knowledge, the early studies were just
confined to isolated system. However, anything can be thought of as
being encompassed by its environment which may influences its
dynamics, so the study of entanglement in an open systems is
imperative. Some inchoate ideas about this topic were presented in
quantum optics \cite{wl73}. On the other hand, with the rise of
relativistic quantum information, much attention has been
concentrated on the behavior of quantum correlations in a
relativistic setting \cite{Ging, Alsing-Mann, Qiyuan, jieci1,
jieci2}. These works provide us some new way in understanding the
quantum theory. Recently, the decoherence in non-inertial frame has
been first discussed under a noise environment \cite{jieci3} also.

It is well known that the Bell state is a concept in quantum
information science and represents the simplest possible examples of
entanglement. And there are four orthogonal Bell states
\begin{eqnarray}\label{Initial-state111}
|\Phi\rangle^{\pm }&=&
(|0\rangle_{A}|0\rangle_{R}\pm |1\rangle_{A}|1\rangle_{R})/\sqrt{2},\nonumber \\
|\Psi\rangle^{\pm }&=&
(|0\rangle_{A}|1\rangle_{R}\pm |1\rangle_{A}|0\rangle_{R})/\sqrt{2},
\end{eqnarray}
where $\{|n\rangle_{A}\}$ indicate Minkowski modes described by
Alice and $\{|n\rangle_{R}\}$ described by Rob, respectively.
Sibasish Ghosh showed that it is not possible to discriminate
between any three Bell states if only a single copy is provided and
if only local operations and classical communication are allowed
\cite{1}. At present most of the studies consider only one of the
Bell states but ignore the other three \cite{Qiyuan, jieci1, jieci2,
jieci3, Salles, Fuentes-Schuller} because different Bell states will
give the same result without considering environment. On the other
hand, Philip Walther and Anton Zeilinger realized a probabilistic
for Bell state analyzer for two photonic quantum bits by use of a
non-destructive controlled-NOT gate based on entirely linear optical
elements \cite{2}. And Miloslav Dusek showed that with no auxiliary
photons it is impossible to discriminate Bell states without errors
and it is impossible to discriminate such Bell states with certainty
in any way by the means of linear optics \cite{3}. Along the way, it
is natural to ask whether the entanglement is related to the initial
(Bell) states if we introduce environment? In this paper, we will
address this question by studying concurrence when both subsystems
are coupled to a noise environment. For the  sake of universality,
we take two general initial states
\begin{eqnarray}\label{Initial-state1111}
|\Theta\rangle_{1}&=&{\sqrt{1-\alpha^2}}|0\rangle_{A}|0\rangle_{R}
+\alpha|1\rangle_{A}|1\rangle_{R}, \\
\label{Initial-state11111}|\Theta\rangle_{2}&=&{\sqrt{1-\alpha^2}}
|0\rangle_{A}|1\rangle_{R}+\alpha|1\rangle_{A}|0\rangle_{R},
\end{eqnarray}
where $-1<\alpha<1,\alpha\neq0$. $|\Theta\rangle_{1}$ can
degrade into the Bell states $|\Phi\rangle^{\pm}$   and $|\Theta\rangle_{2}$  into  $|\Psi\rangle^{\pm}$ if
we take $\alpha=\pm1/\sqrt{2}$, respectively.  Then, we can find that the behavior of the entanglement will be greatly influenced by
initial states, but we can only distinguish the initial states
$|\Theta\rangle_{1}$ (or $|\Phi\rangle^{\pm}$) from
$|\Theta\rangle_{2}$ (or $|\Psi\rangle^{\pm}$).

In this paper, we will investigate the dependence of the entanglement on the initial states which reduce to four orthogonal Bell states under amplitude damping channel. We will show that the entanglements for different initial states will decay along different curves even with the same acceleration and parameter of the states, and the possible range of the sudden death of the entanglement for 1 is larger than that for 2.

This paper is structured as follows. In Sec. II we will study the
concurrence when both of the qubits under amplitude damping channel
using the initial state $|\Theta\rangle_{1}$. In Sec. III we will
consider the concurrence when both of the qubits under the same
environment by taking the state $|\Theta\rangle_{2}$. Our work will
be summarized in last section.

 \section{Entanglement for  initial  states $|\Theta\rangle_{1}$}

\vspace*{0.2cm} We first study the entanglement for  initial  states
$|\Theta\rangle_{1}$. We assume two observers, Alice who stays
stationary has a detector only sensitive to mode $|n\rangle_{A}$ and
Rob who moves with a uniform acceleration has a detector which can
only detect mode $|n\rangle_{R}$, share a entangled initial state
$|\Theta\rangle_{1}$ at the same point in Minkowski spacetime.  We
can use a two-mode squeezed state to expend the Minkowski vacuum
from the perspective of Rob \cite{Alsing-Mann}  $ |0\rangle_{M}=
\cos r|0\rangle_{I}|0\rangle _{II}+\sin r|1\rangle_{I}|1\rangle
_{II},$ where $\cos r=(e^{-2\pi\omega c/a}+1)^{-1/2}$,  $a$ is Rob's
acceleration, $\omega$ is energy of the Dirac particle,  $c$ is the
speed of light in vacuum, and $\{|n\rangle_{I}\}$  indicate Rindler
modes in region $I$ and $\{|n\rangle_{II}\}$ indicate Rindler modes
in region $II$, respectively. And the only excited state can be
given by $ |1\rangle_{M}=|1\rangle_{I}|0\rangle_{II}. $ Thus, we can
rewrite Eq. (\ref{Initial-state1111}) in terms of Minkowski modes
for Alice and Rindler modes for Rob
\begin{eqnarray} \label{state}
|\Theta\rangle_{A, I, II}&=&{\sqrt{(1-\alpha^2)}} \cos
r|0\rangle_{A}
|0\rangle_{I}|0\rangle_{II}+\alpha|1\rangle_{A}|1\rangle_{I}|0\rangle_{II}+\nonumber
\\&&{\sqrt{(1-\alpha^2)}}\sin r|0\rangle_{A}
|1\rangle_{I}|1\rangle_{II}.
\end{eqnarray}
On account of Rob is causally disconnected from region  $II$, and
tracing over the states in region $II$, we obtain
\begin{eqnarray}\label{eq:state1E}
 &&\rho_{1}=\left(
  \begin{array}{cccc}
    (1-\alpha^2)\cos^2 r& 0 & 0 &\alpha{\sqrt{(1-\alpha^2)}} \\
    0 &(1-\alpha^2)\sin^2 r & 0 & 0 \\
    0 & 0 & 0& 0 \\
   \alpha{\sqrt{(1-\alpha^2)}} & 0 & 0 &\alpha{\sqrt{(1-\alpha^2)}} \\
  \end{array}
\right).\nonumber \\
\end{eqnarray}
We now let both Rob and Alice interact with a amplitude damping
environment \cite{Brune1}. There is a simple way to understand this
process if we use the quantum map \cite{Breuer}
\begin{eqnarray}
\label{AmplitudeDampingMap}
|0\rangle_{S}|0\rangle_E&\rightarrow&
|0\rangle_{S}|0\rangle_E  \label{en1}\; , \\
|1\rangle_{S}|0\rangle_E&\rightarrow&
\sqrt{1-P}|1\rangle_{S}|0\rangle_E +
\sqrt{P}|0\rangle_{S}|1\rangle_E  \label{en2}\; .
\end{eqnarray}
Eq. (\ref{en1}) shows that if the system stays $|0\rangle_{S}$ both
it and its environment will not change at all.  Eq. (\ref{en2})
indicates that if the system stays $|1\rangle_{S}$ the decay will
exist in the system with probability $P$,  and it can also remain
there with probability $(1-P)$.

If the environment  acts  independently on Alice's and Rob's states,
the total evolution of these two qubits system can be expressed as
\cite{Salles} $ L(\rho_{AR})=\sum_{\mu \nu} M^{A}_\mu \otimes
M^{R}_\nu \rho_{AR} M_\nu^{A\dag}\otimes M_\mu^{R\dag},  $ where
$M_{\mu}^{i}$ are the Kraus operators
\begin{eqnarray}
M_0^{i}=\left(\begin{array}{cc}
           1&0\\
           0&\sqrt{1-P_{i}}
           \end{array}\right),  &\; & M^{i}_1=\left(\begin{array}{cc}
                                           0&\sqrt{P_{i}}\\
                                           0&0
                                          \end{array}\right),
\label{Kraus1A}
\end{eqnarray}
where  $i=(A, ~R)$, $P_{A}$ is the decay parameter in Alice's quantum
channel and $P_{R}$ is the decay parameter in Rob's quantum channel, and $P_i$ ($0\leq P_i\leq1$) is a parameter relating
only to time.   Under the Markov approximation,  the relationship
between the parameter $P_i$ and the time $t$ is given by
$P_i=(1-e^{-\Gamma_i t})$ \cite{Brune1, Salles}, where $\Gamma_i$ is the
decay rate. We must note that here we just consider the local
channels \cite{Salles}, in which all the subsystems interact
independently with its own environment and no communication appears.
i.e.,  $P_{A}=P_{R}=P$. Then we can obtain the evolved states in this case
\begin{widetext}
 \begin{eqnarray}\label{eq:state1E1}
 &&\rho_{s1}=\left(
  \begin{array}{cccc}
    P^2\alpha^2+\gamma(\cos^2+P\sin^2 r) & 0 & 0 &\alpha \beta\sqrt{\gamma}\cos r \\
    0 &\beta(P\alpha^2+\gamma\sin^2 r) & 0 & 0 \\
    0 & 0 & P\beta\alpha^2& 0 \\
    \alpha\beta\sqrt{\gamma}\cos r & 0 & 0 &\beta^2\alpha^2  \\
  \end{array}
\right),
\end{eqnarray}
\end{widetext}
where $\beta=1-P$ and $\gamma=1-\alpha^2$. Since it is well known
that the degree of entanglement for a two-qubits mixed state in
noisy environments can be quantified very  conveniently by the
concurrence \cite{Wootters, Coffman} $ C_{s} =\max \left\{ 0,
\sqrt{\lambda _{1}}-\sqrt{\lambda _{2}}-\sqrt{\lambda
_{3}}-\sqrt{\lambda _{4}}\right\},  \quad\lambda_i\ge
\lambda_{i+1}\ge 0, $ where $\sqrt{\lambda_i}$ are square roots of
the eigenvalues of the matrix $\rho_{s}\tilde{\rho}_{s}$, with
$\tilde{\rho}_{s}=(\sigma_y\otimes\sigma_y)\, \rho_{s}^{*}\,
(\sigma_y\otimes\sigma_y)$ is the ``spin-flip" matrix for the state
(\ref{eq:state1E}). So, we obtain the concurrence as a function of
$\alpha$, $r$ and $P$
\begin{eqnarray} \label{concurrence1}
C_{s1}&=&2|\alpha|(1-P)\Big\{\sqrt{1-\alpha^2} \cos
r\nonumber \\ & &-{\sqrt{P[P\alpha^2+(1-\alpha^2)\sin^2 r]}}\Big\}.
\end{eqnarray}
Due to the concurrence is just depended on $\alpha^2$ and
$|\alpha|$,  we can't distinguish the initial states described by
$|\Theta\rangle_{1}$ with $1>\alpha>0$ or $-1<\alpha<0$.

\section{Entanglement for initial  states $|\Theta\rangle_{2}$}

\vspace*{0.2cm} Now, we consider the other initial state
$|\Theta\rangle_{2}$. Using the same method as mentioned above  we
obtain its density matrix
\begin{eqnarray}\label{eq:state2E}
&&\rho_{2}=\left(
  \begin{array}{cccc}
    0 & 0 & 0 & 0\\
    0 &1-\alpha^2 & \alpha\sqrt{1-\alpha^2}\cos r & 0\\
    0 & \alpha\sqrt{1-\alpha^2}\cos r & \alpha^2\cos^2 r & 0\\
    0 & 0 & 0 &\alpha^2\sin^2 r\\
  \end{array}
\right), \nonumber \\
\end{eqnarray} and the evolved state for $|\Theta\rangle_{2}$
\begin{widetext}
\begin{eqnarray}\label{eq:state2}
\rho_{s2}=\left(
\begin{array}{cccc}
    P\gamma+P\alpha^2(\cos^2 r+P\sin^2 r) & 0 & 0 & 0\\
    0 &\beta(\gamma+P\alpha^2\sin^2 r) & \alpha \beta\sqrt{\gamma}\cos r & 0 \\
    0 & \alpha \beta\sqrt{\gamma}\cos r  & \beta\alpha^2(\cos^2 r+P\sin^2 r) & 0 \\
    0 & 0 & 0 &\beta^2\alpha^2\sin^2 r\\
  \end{array}
\right).
\end{eqnarray}
\end{widetext}
Thus, the concurrence is
\begin{eqnarray} \label{concurrence2}
C_{s2}&=&2|\alpha|(1-P)\Big\{\sqrt{1-\alpha^2} \cos r\nonumber
\\ &-&\sin r \sqrt{P[(1-\alpha^2)+\alpha^2(\cos^2 r +P \sin^2
r)]}\Bigg\}.
\end{eqnarray}
From which we know that we can't distinguish the initial states
described by $|\Theta\rangle_{2}$ with $1>\alpha>0$ or
$-1<\alpha<0$, too.

\section{Discussions and Conclusions}

By comparing Eqs. (\ref{concurrence1}) and (\ref{concurrence2}), we
can see that there are obvious differences between $C_{s1}$ and
$C_{s2}$. Especially, we find that $C_{s1}=(1-P)^2$ and
$C_{s2}=(1-P)$ for Bell states  ($\alpha=1/\sqrt{2}$) in an inertial
frame. But if $P=0$, we have $C_{s1}=C_{s2}$ for any $r$ and
$\alpha$,  which means that the two groups of the initial states
will be equivalent without the effect  of environment.

\begin{widetext}

\begin{figure}[ht!]
  \centering
\includegraphics[scale=0.425]{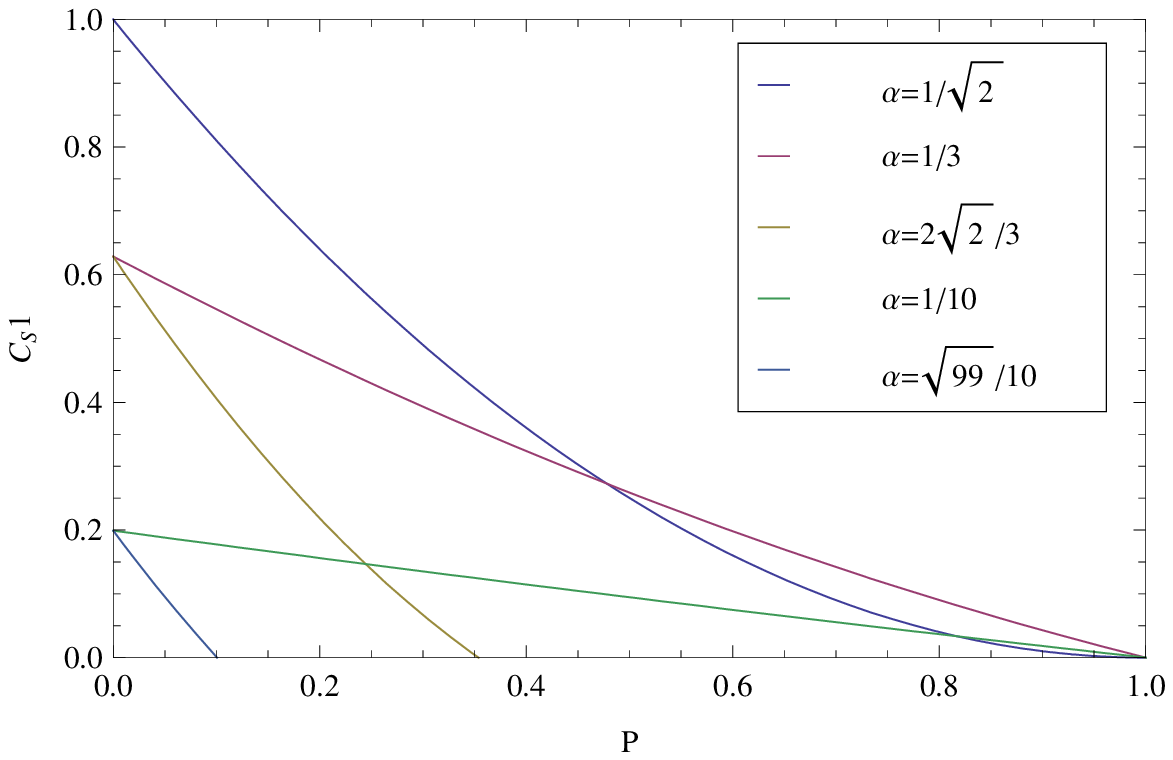}
\includegraphics[scale=0.425]{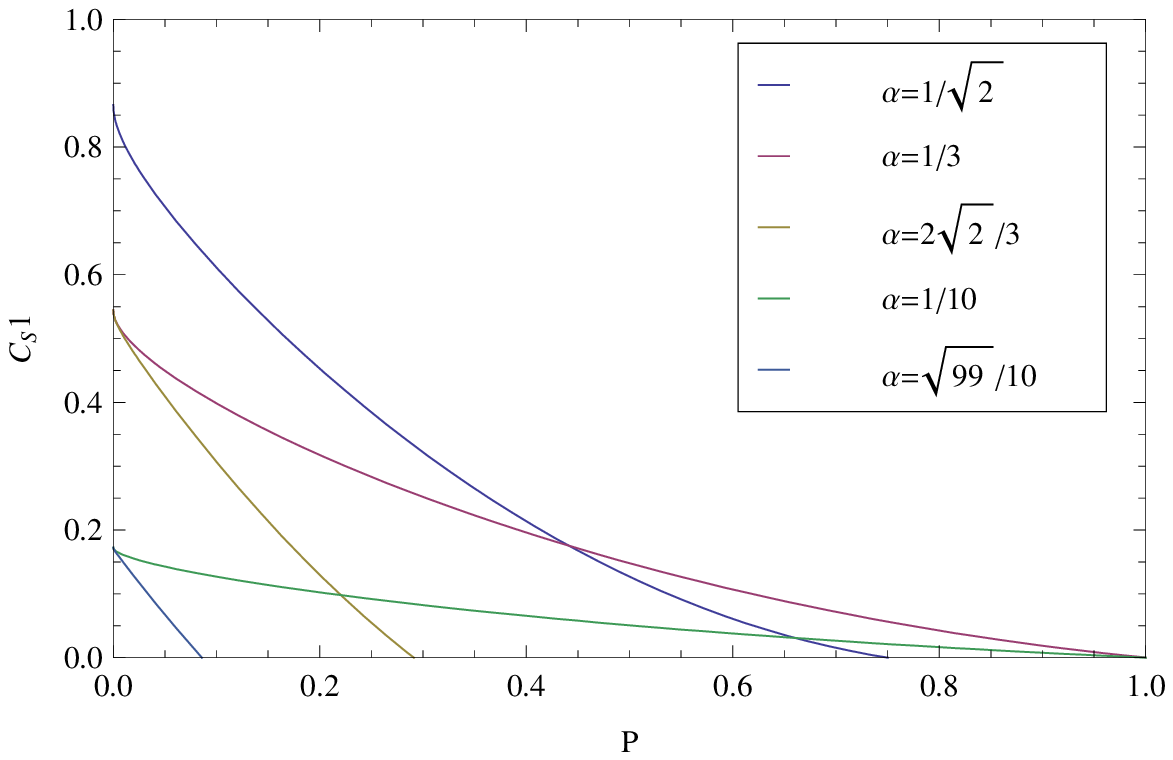}
\includegraphics[scale=0.425]{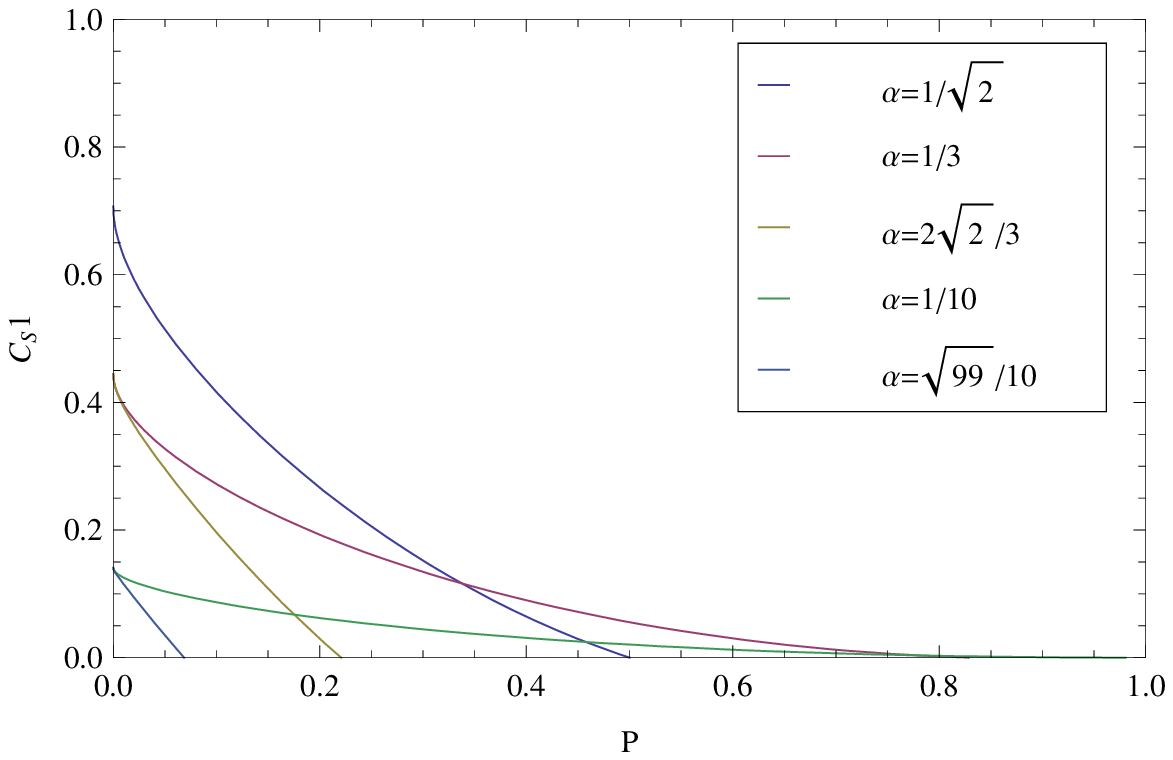}
\\
\includegraphics[scale=0.425]{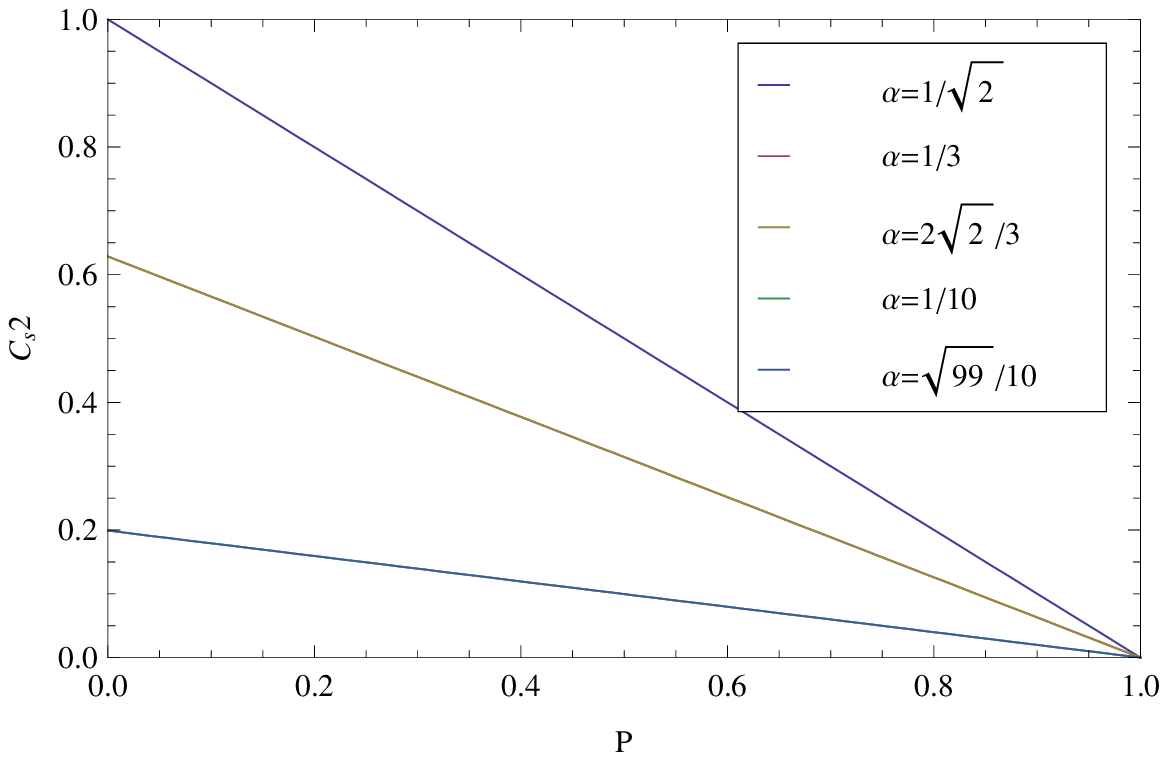}
\includegraphics[scale=0.425]{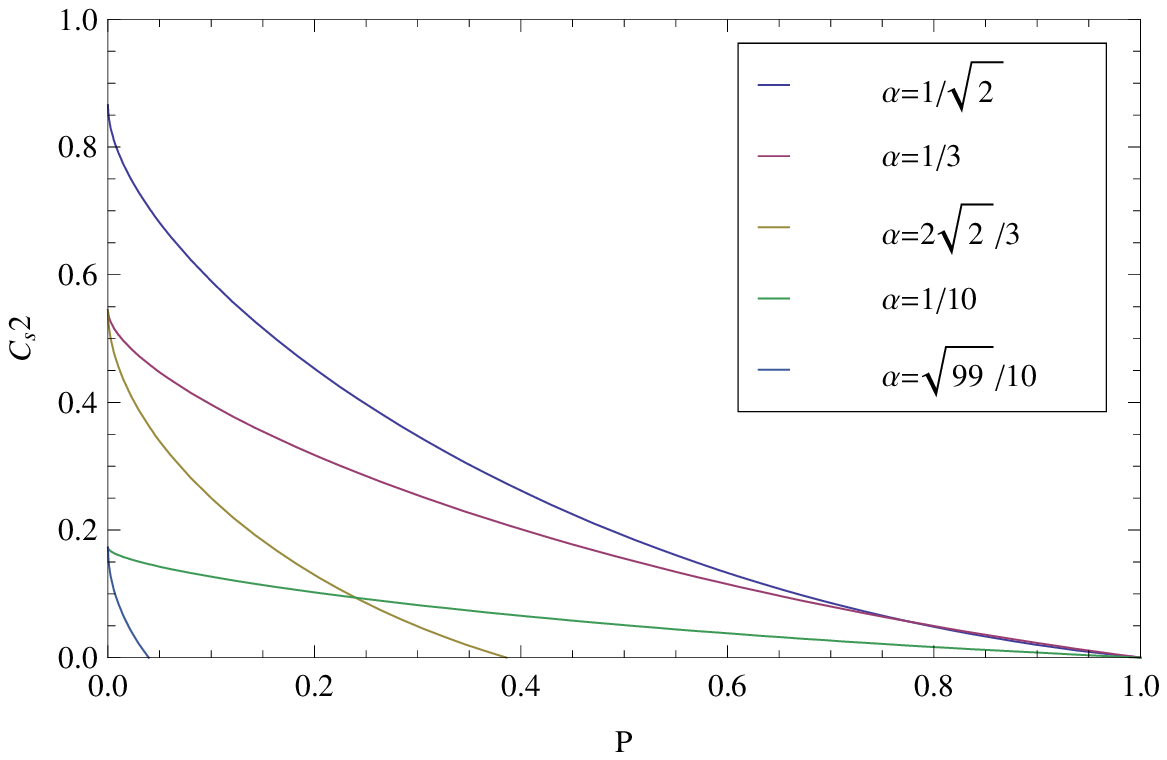}
\includegraphics[scale=0.425]{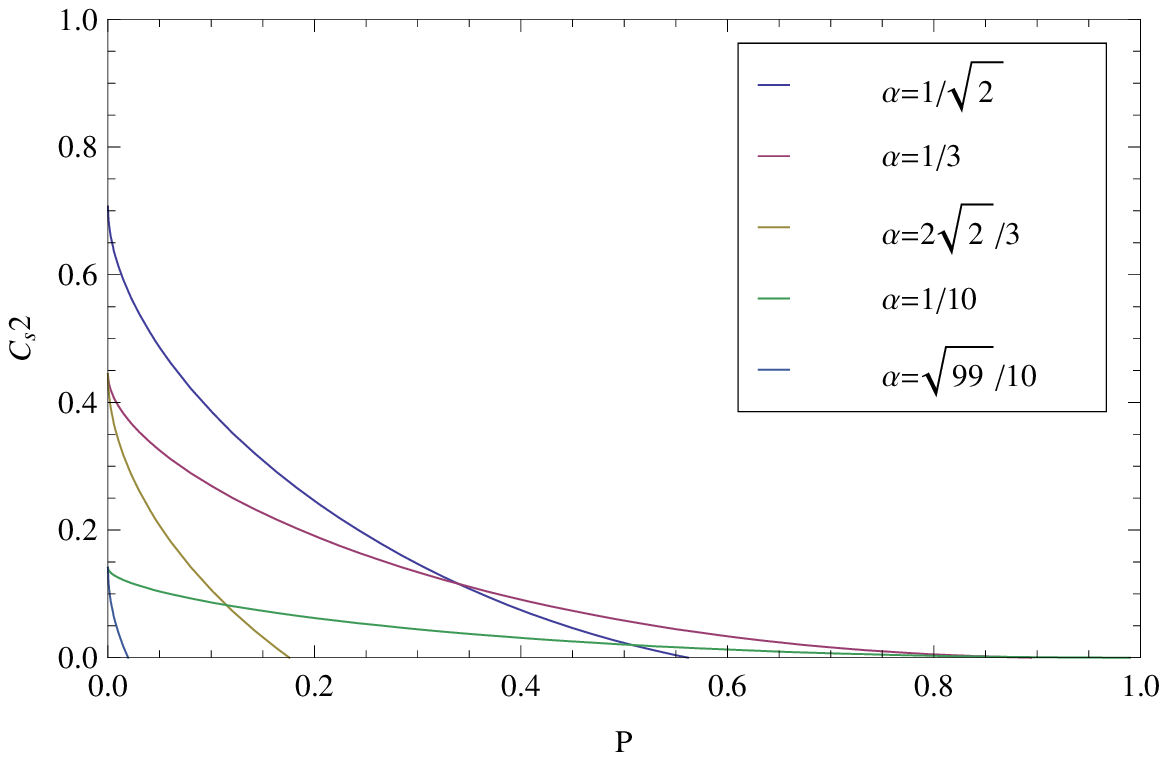}
\caption{\label{B1}(Color online) Concurrence for the initial states
$|\Theta\rangle_{1}$ (first row) and $|\Theta\rangle_{2}$ (second row)
as a function of $P$ with some fixed initial state parameters
$\alpha$ and acceleration parameters $r=0$ (left),    $r=\pi/6$
(middle) and $r=\pi/4$ (right)
 when both qubits are coupled to the same noise environment.  }
\end{figure}
\end{widetext}

To learn the behavior of the entanglement intuitively, we plot the concurrence for different initial states
$|\Theta\rangle_{1}$ and $|\Theta\rangle_{2}$ with different
parameters in Fig. \ref{B1}. From the left two panels we find that, in an
inertial frame (i.e., $r=0$), the $C_{s1}$ will tend to  zero for a finite time which is called sudden death if $\alpha>1/\sqrt{2}$.
However, the $C_{s2}$ will not tend to zero for any $\alpha$ and
it will decay along the same curve for both $\alpha$ and its ¡°normalized partner¡± $\sqrt{1-\alpha^2}$, which
shows us that we can't discriminate Alice's excited states from Rob's excited states for initial states $|\Theta\rangle_{2}$, i.e.,  $\alpha$  and $\sqrt{1-\alpha^2}$ will lead to a symmetrical structure at $r=0$ for initial states $|\Theta\rangle_{2}$.
We also note that the concurrences for $|\Theta\rangle_{1}$
and $|\Theta\rangle_{2}$ decay different from each other even they
have the same $\alpha$.
From Fig. \ref{B1}, if we fix $\alpha$, it is easy to find out that,
as $r$ becomes large which means the increase of the Rob's
acceleration, the sudden death of the entanglement for both
$|\Theta\rangle_{1}$ and $|\Theta\rangle_{2}$ would happen earlier
and earlier. That is to say, a bigger acceleration leads to a faster
decay of the entanglement, in another word, the stronger Unruh
effect  will  speed the decay of entanglement. On the other hand, if
we fix $r$, we  find that the entanglement decay faster and faster
as the $\alpha$ increases except a special case for
$|\Theta\rangle_{2}$ with $r=0$.  For the states
$|\Theta\rangle_{1}$, the more the initial excited states there are,
the stronger is the interaction between the system with environment,
which will lead to a faster disappear of the entanglement. For the
states $|\Theta\rangle_{2}$, although the total number of the
excited states keeps conservable whatever $\alpha$ is, the time of
sudden death can also change with $\alpha$ because the proportion of
Alice's excited states and Rob's excited states affects the decay
velocity.

\begin{figure}[ht!]
\includegraphics[scale=0.65]{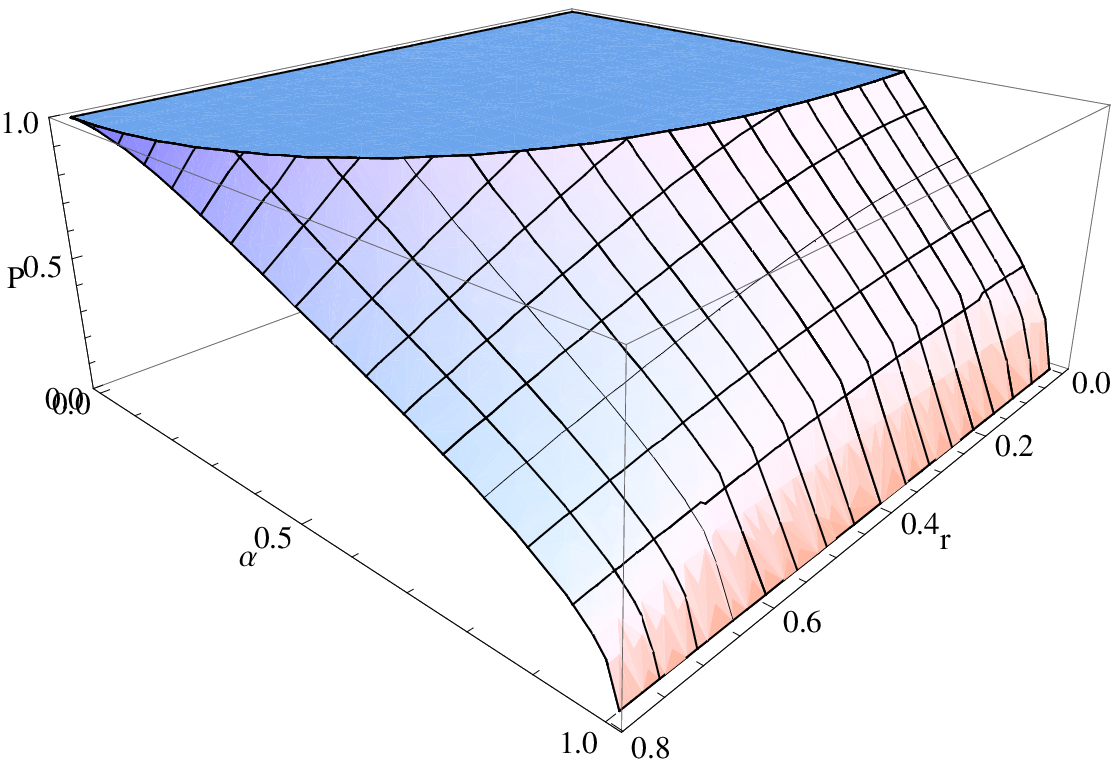}
\includegraphics[scale=0.65]{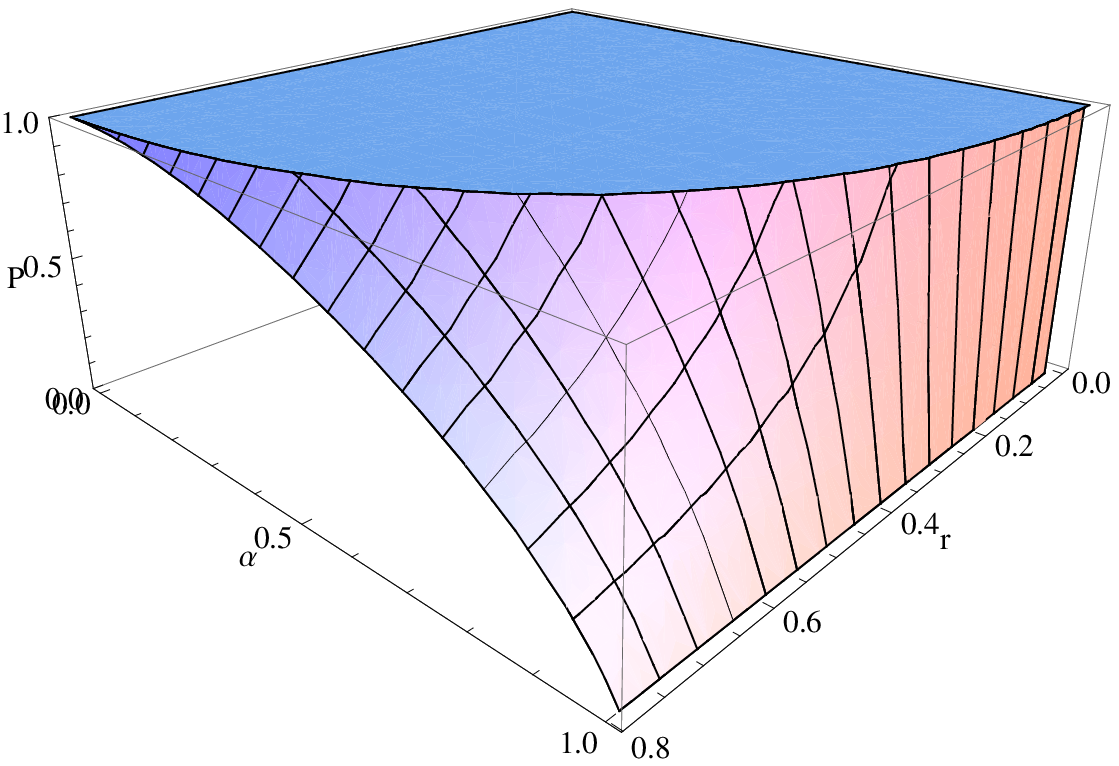}
\caption{\label{B4}(Color online) The grid surface presents the
possible range of the sudden death for the initial states
$|\Theta\rangle_{1}$ (left) and $|\Theta\rangle_{2}$ (right) when
both qubits are coupled to a noise environment.}
\end{figure}

If  the parameters $r$, $\alpha$ and $P$  in Eq. (
\ref{concurrence1}) satisfy the relation
\begin{eqnarray}\label{cs11}
|\alpha|=\sqrt{\frac{1-P+\cos 2r +P\cos 2r}{1-P+2P^2+\cos 2r+P\cos
2r}},
\end{eqnarray}
we have $C_{s1}=0$,
and if the parameters $r$, $\alpha$ and $P$ in  Eq. (\ref{concurrence2}) meet
\begin{eqnarray}\label{cs22}
|\alpha|=2\sqrt{\frac{1-P+\cos 2r +P\cos 2r}{1+4P+4(1+P-P^2)\cos
2r+(P-1)P\cos 4r}},\nonumber \\
\end{eqnarray}
we obtain $C_{s2}=0$. Using Eqs. (\ref{cs11}) and (\ref{cs22}) (See
Fig. \ref{B4}), we can find a possible range for the sudden death of
the entanglement. In consideration of $0<P<1$, for the states
$|\Theta\rangle_{1}$, we find that the sudden death of the
entanglement will appear if $\alpha$ satisfy the relation
\begin{eqnarray}
1>|\alpha|>\frac{\sqrt{cos 2r}}{\sqrt{1+\cos 2r}}. \end{eqnarray} And
for the states $|\Theta\rangle_{2}$, the sudden death of
entanglement can happen only when
\begin{eqnarray}
1>|\alpha|>\frac{\sqrt{cos 2r}}{\cos r}. \end{eqnarray} It is
obviously that the possible range of the sudden death of the entanglement for
$|\Theta\rangle_{1}$ is larger than that for $|\Theta\rangle_{2}$.
If $\alpha<\sqrt{3}/2$, whatever $r$ is, the disappear of the entanglement for $|\Theta\rangle_{1}$ will be earlier than that for $|\Theta\rangle_{2}$.

Above discussions reveal some different behaviors of concurrences
for the initial states $|\Theta\rangle_{1}$ and $|\Theta\rangle_{2}$
when both subsystems are coupled to noise environment. Thus, the
entanglement is dependent to the initial states under the amplitude
damping channel.

\begin{acknowledgments}
\vspace*{0.50cm}

This work was supported by the National Natural Science Foundation of China under Grant Nos. 11175065, 10935013; the SRFDP under Grant
No. 20114306110003; PCSIRT, No. IRT0964; the Hunan Provincial Natural Science Foundation of China under Grant No. 11JJ7001;  and the Construct Program of the National  Key Discipline.

\end{acknowledgments}

\vspace*{0.2cm}

%\end{acknowledgments}

\end{document}